\documentclass[12pt,preprint]{aastex}
\def\lax {\ifmmode{_<\atop^{\sim}}\else{${_<\atop^{\sim}}$}\fi}  
\def\gax {\ifmmode{_>\atop^{\sim}}\else{${_>\atop^{\sim}}$}\fi}  
\def\gtorder{\mathrel{\raise.3ex\hbox{$>$}\mkern-14mu
             \lower0.6ex\hbox{$\sim$}}}
 
\begin{document}

\title{Discovery of Red-Skewed K$_\alpha$ iron line  
in Cyg~X-2 with {\it Suzaku}}

\author{ Nikolai Shaposhnikov\altaffilmark{1,2}, Lev Titarchuk\altaffilmark{3} and Philippe Laurent\altaffilmark{4}}
 
\altaffiltext{1}{ CRESST/NASA GSFC, Astrophysics Science
 Division, Greenbelt MD 20771; nikolai@milkyway.gsfc.nasa.gov}
\altaffiltext{2}{University of Maryland, Astronomy Department, College Park, MD 20742}
\altaffiltext{3}{George Mason University/Center for Earth Observing
and Space Research, Fairfax, VA 22030; US Naval Research Laboratory,
Code 7655, Washington, DC 20375; email: Lev.Titarchuk@nrl.navy.mil; 
 NASA GSFC, code 661, 
Greenbelt MD 20771, USA; email:lev@milkyway.gsfc.nasa.gov}
\altaffiltext{4}{CEA/DSM/DAPNIA/SAp, CEA Saclay, 91191 Gif sur Yvette, France;plaurent@cea.fr; Laboratoire APC, 10 rue Alice Domont et Leonie Duquet, 75205 Paris Cedex 13, France}

\begin{abstract}
We report on the {\it Suzaku} observation of neutron star low-mass X-ray binary 
Cygnus X-2 which reveals  a presence of the iron K$_\alpha$ emission line.
The line profile shows a significant red wing. This discovery 
increases the number of neutron star sources where red-skewed iron lines were
observed and strongly suggests that this phenomenon is common not only 
in black holes but also in other types of accreting compact objects. We examine the line 
profile in terms of models which attribute its production to the 
relativistic effects due to  reflection of X-ray radiation from a cold accretion disk
and also  as a result of the line formation in the extended wind/outflow configuration.
Both models are able to adequately represent the observed line profile.
We consider the results of line modeling in the context of subsecond 
variability. While we were unable to conclusively disqualify one of the 
models, we find that the wind paradigm has several advantages over the relativistic
disk reflection model.

\end{abstract}

\keywords{accretion, accretion disks ----stars: neutron
---X-rays: individual (Cygnus X-2) --- stars:}

\section{Introduction}

Recent discoveries of red-skewed iron lines in spectra
of neutron star (NS) sources Serpens X-1 \citep{bs07}, 4U 1820-30, GX 349+2 
\citep[][C08 hereafter]{cack08} and 4U 1636-536 \citep{pan08}
show that the phenomenon of red-skewed lines is not restricted to black hole 
(BH) sources. In this Paper we report on the discovery of the asymmetric 
iron line  in the  {\it Suzaku} spectrum of the NS source Cygnus X-2 (Cyg X-2).
Therefore, Cyg X-2 is the fifth NS source which shows strongly asymmetric iron line profile. This 
indicates that the red-skewed lines in NS sources may be as common as in BH sources.
More generally, asymmetric emission lines appear to be abundant in both types
of accretion powered X-ray sources.
It is crucial to correctly identify the physical origin of the red-skewness of
these lines because they can be potentially used to study the properties of the
accretion close to accreting objects as well as to constrain the fundamental
characteristics of compact objects.

C08 interpreted the $K_\alpha$ iron line profiles 
in terms of relativistically red-shifted emission due to reflection off the
accretion disk very close to a compact object. This scenario is commonly
accepted as an explanation of strongly red-skewed iron lines in BH sources 
\citep{mil07}. The main motivation for
applying the relativistic line formation scenario to the NS case
was the fact that the inner radius of the 
accretion disk, which was predicted by the relativistic line model was consistent with 
the interpretation of the 
highest observed kilohertz quasi-periodic oscillation (kHz QPO) frequency in 
this sources as a Keplerian frequency at this radius. 

A red-skewed profile of emission lines can be  also produced 
by repeated electron scattering in a diverging outflow as proposed by 
\citet{lt07}, LT07 hereafter, see also references therein.
In the framework of the wind model  the fluorescent iron line K$_{\alpha}$  is formed 
in the partly ionized wind as a result of illumination by central source. 
Electron scattering of the iron K$_\alpha$ photons within the ionized expanding flow 
leads to a decrease of their energy (redshift).
This photon redshift is an intrinsic property of any
outflow for which divergence is positive. Recently \citet{sim08} confirmed
using multidimensional Monte-Carlo simulation that for sufficiently high wind densities, moderate 
Fe K$_{\alpha}$ emission lines can be formed  and that electron scattering in the flow may cause these lines 
to develop extended red wings.

We examine the red-skewed line profile observed in Cyg X-2 both in terms of  
the relativistic paradigm and in the framework of the wind downscattering. 
The main  obstacle in analyzing  the high spectral resolution data 
with the wind model is that  analytical solution is
not available for the general formulation of this problem. 
LT07  used Monte Carlo (MC) simulations 
to model the line profiles produced in the wind environment. 
In the presented work we provide a consistent analysis of the line profile with 
the wind model by introducing the LT07 MC code into XSPEC astrophysical data analysis 
package. We find that the wind
outflow model is able to reproduce the red-skewed line profile 
with the fit quality similar to that shown by the relativistic reflection models.
Therefore, in order to distinguish between these two models one has to consider 
their consistency in a broader phenomenological context. For example, 
C08 proposed to look for high frequency quasi-periodic oscillations (HF QPOs)
as an additional evidence of a Keplerian disk existence close to a NS, which then
would necessitate the presence of the reflection components in energy spectra.
In contrast, the presence of the opaque wind in the system would result
in smearing the signal coming from the central region. 
This smearing effect leads to a suppression of  the fast variability of X-ray emission.
Unfortunately, high time resolution observations  by {\it Rossi X-ray Timing Explorer} ({\it RXTE}),
simultaneous  with the  {\it Suzaku}   observations analyzed in this Paper,  are not 
available. Thus, we resort to a study based on the {\it RXTE}
data set with similar spectral characteristics. Our comparative analysis indicates that 
the source was in low variability state during these observations which is more consistent
with the wind/outflow scenario.



Description of {\it Suzaku} and supporting {\it RXTE} observations 
as well as details of our spectral fitting are presented are presented in \S 2. 
We discuss implications of the relativistic and wind outflow red-skewed line formation scenarios
in Cyg X-2 in \S 3. Conclusions follow in \S 4.

\section{Observations and Spectral Modeling}
\label{observ}

Cyg~X-2 is a low mass X-ray binary \citep[see][for review]{lew95} which
 exhibits a Z-shape color-color diagram \citep{hvdk89}.
The observations of thermonuclear X-ray bursts \citep{s98}
identified the nature of the compact object in Cyg X-2 as
a neutron star. \citet{ts02} used the {\it RXTE} burst data 
to estimate the NS mass to be about 1.4 solar masses and radius to be about 9 km. 
\citet{wij98} reported the simultaneous detection of twin kHz peaks at
500 and 860 Hz and highest single kHz QPO at 1007 Hz.

Cyg X-2 was observed by {\it Suzaku} \citep{mit07} on May 16, 2006 for a total 
exposure of 39 ksec (Observation ID: 401049010). However, during more than a half of the
observation the satellite were operating in the medium telemetry mode which led 
to the telemetry saturation and resulted in data unusable for scientific analysis.
During intervals when  the high telemetry setting were utilized
the foreground illuminated  XIS detectors (i.e. XIS 0, 2 and 3) 
operated in  3$\times$3 Event Editing mode with Burst Clock and 1/4 Window settings.
The 3$\times$3 Event Editing mode is not available for the XIS 1 for this observation.
We therefore used data from XIS 0, 2, and 3 collected when high telemetry rate was utilized.
We reduce the {\it Suzaku} XIS and HXD data using the {\it xselect} data analysis tool
following the guidelines given by {\it Suzaku} Data 
Analysis Guide \footnote{http://heasarc.gsfc.nasa.gov/docs/suzaku/aehp\_data\_analysis.html}. 
The XIS images indicate that strong pile-up in the center of the source point spread function (PSF) 
leads to a  characteristic ``crater''. In order to remove piled-up data we  extracted 
XIS spectra from the annulus regions with the outer radius set to the maximum allowed
by the XIS detector field of view ($\sim 115$ arcsec) and the inner radius was
manually selected to excise the most piled-up inner core of PSF ($\sim 15$ arcsec).
The spectra and corresponding responses were extracted by {\it xselect} 
extractor and {\it xisresp},  {\it xisrmfgen} tools. Spectra and responses 
for individual detectors were then added using {\it mathpha} and {\it addrmf}  FTOOLS.
We linearly rebin XIS spectral and response data to obtain 1024 spectral channels.
HXD/PIN spectrum was corrected for non-X-ray and cosmic X-ray background.
We fit XIS and PIN spectra jointly in XSPEC using 0.7-9.0 keV energy range for XIS 
data and 15.0-40.0 keV range for PIN data fixing the cross-normalization factor 
between XIS and PIN spectra at unity. Due to large calibration 
uncertainties we also ignore 1.5-2.5 keV range for XIS spectrum.

\begin{figure}
\includegraphics[scale=0.33,angle=-90]{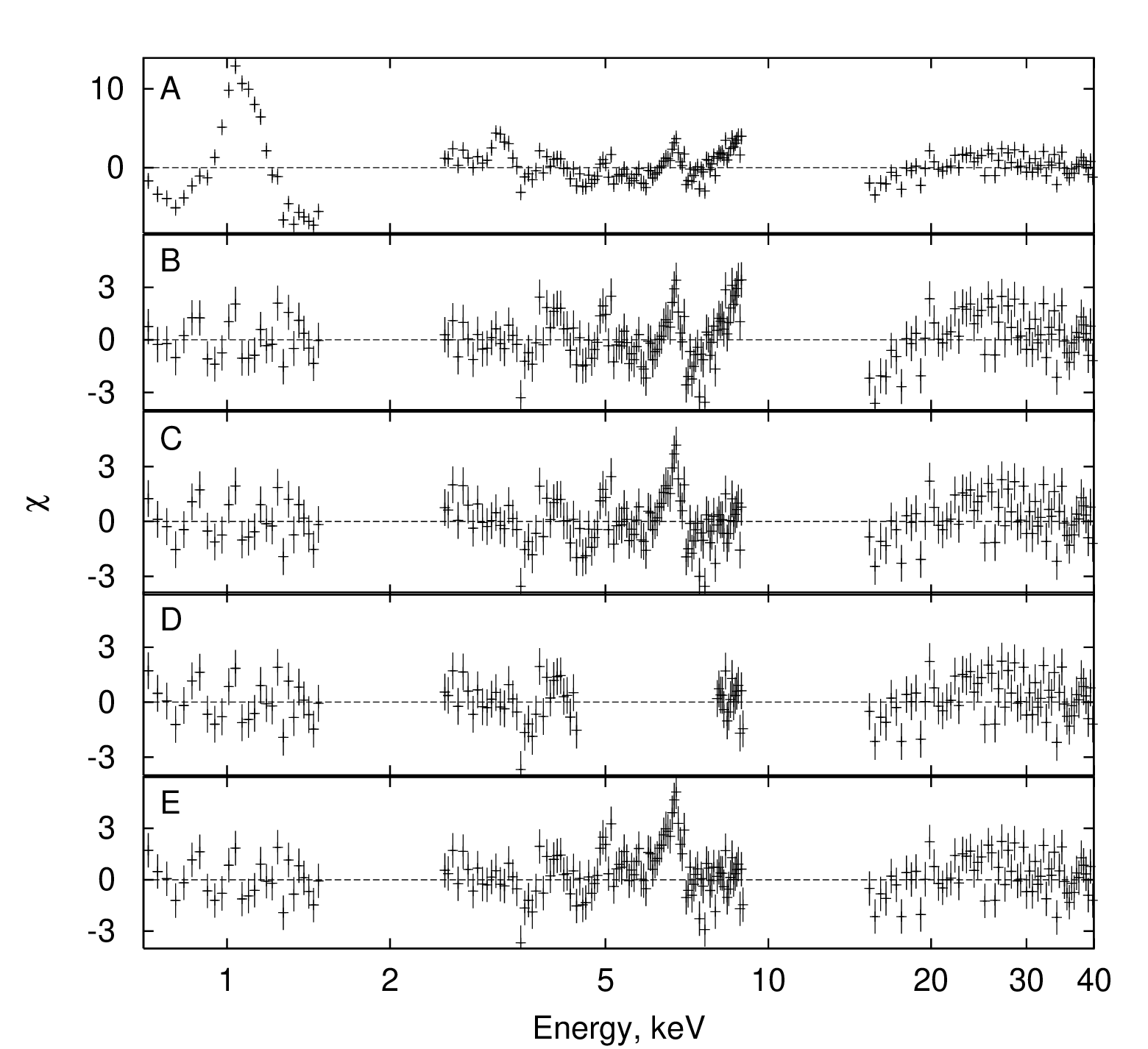}
\includegraphics[scale=0.33,angle=-90]{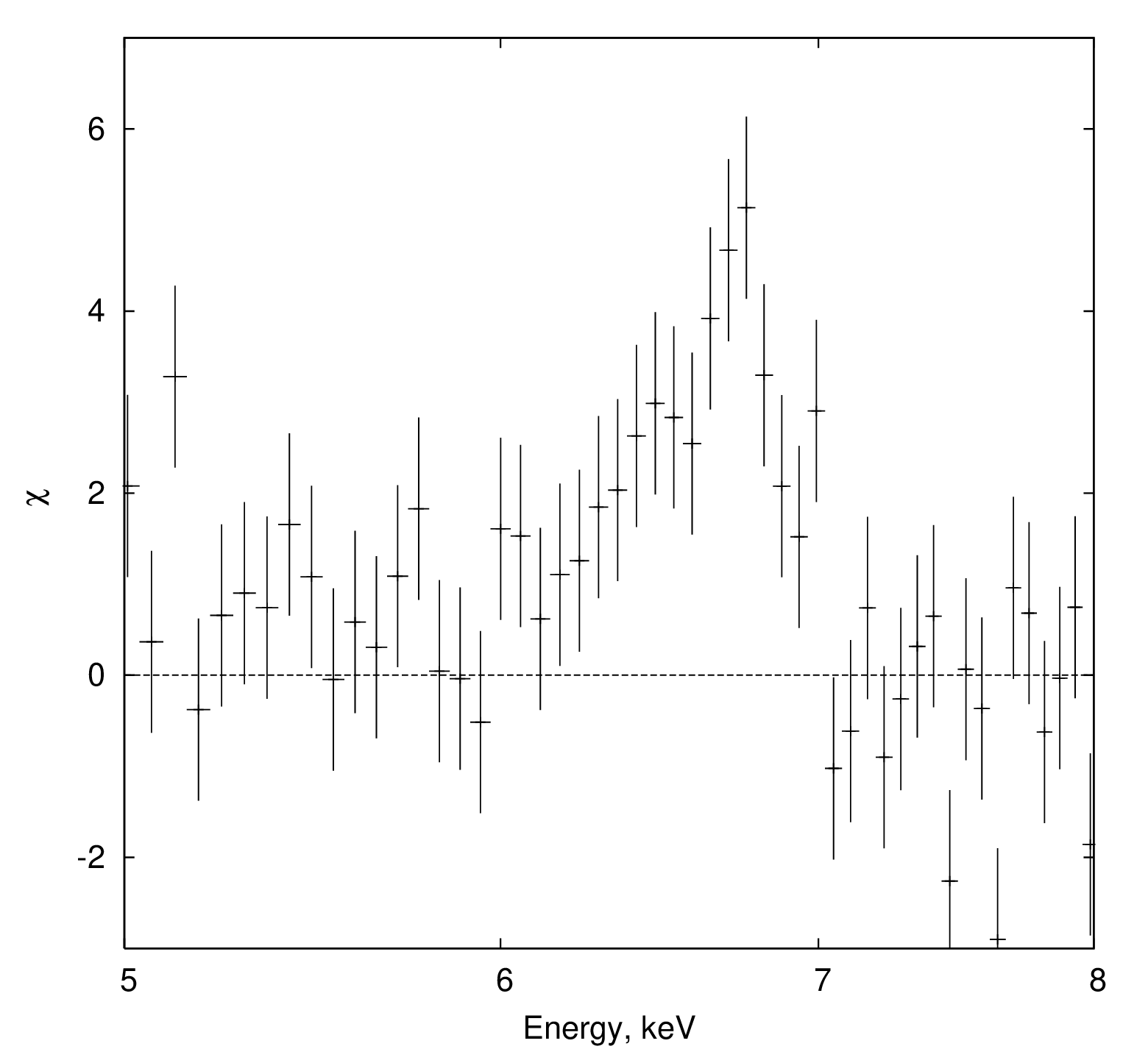}
\caption{ {\it Left panel:} Identification of the continuum model and narrow lines in the {\it Suzaku} spectrum of Cygnus X-2. Model residuals in units of one sigma error in corresponding channel are shown
for  A) fit with continuum model {\tt wabs(comptt+bb)} only ($\chi^2_{red}=7.5$). B) fit with lines added at 1 keV and 3.2 keV [{\tt wabs(gaussian+gaussian+comptt+bb)}, $\chi^2_{red}=2.0$]. C) file up model includes [{\tt pileup*wabs(gaussian+gaussian+comptt+bb)}, $\chi^2_{red}=1.7$] .  D) fit with energy range from 4.5 keV to 7.5 keV excluded and E) the model obtained in D with channels between 4.5 keV and 7.5 keV noticed. {\it Right panel:} Panel D on the right side zoomed in the iron line region. Emission line with an apparent red-skewness is seen in the data.}
\label{line_res}
\end{figure}

For the continuum spectra we choose the sum 
of thermal ({\tt bbody}) and Comptonized  
\citep[{\tt comptt};][]{comptt} components, modified by interstellar photoelectric absorption 
according to \citet{wabs}. When we directly fit the data with this model 
we observe three distinct narrow features in the residuals (See Figure \ref{line_res}, left 
panel A). First, we see a line signature at 6-7 keV, which is the primary target of our
investigation. We also observe a weak excess around 3.2 keV and
a prominent line at 1 keV. 
The feature at 3.2 keV is  probably an instrumental artifact. 
The line at 1 keV was reported previously from Cyg X-2 \citep{s93} and presumably  
belongs to the source spectrum. Both features are well represented by {\tt gaussian} shape
(see residuals on  panel B of Figure 1).
In addition  to these narrow lines we observe a building-up excess towards
higher energies in XIS spectrum. This indicates a presence of a residual 
pile-up in the regions close to the excised central part of the PSF.
To mitigate this effect we utilize the XSPEC {\tt pileup} convolution model
designed to model pile-up effect in CCD detectors. 
The XSPEC implementation of the pile-up model was initially designed 
to describe this effect in {\it Chandra} data. However, the model
, developed in  \citet{pileup}, is valid for {\it Suzaku} XIS detectors also.
{\it Suzaku}  PSF of XIS CCDs is broader and spreads around larger number of
pixels. Therefore, pile-up model for the XIS spectra would require larger 
number of regions to be considered for pile-up. 
To model pile-up in XIS detectors we used the following  parameter settings for the {\it pileup} XSPEC model: 
2 second time frame, maximum of 10 photons to pile up, unity for grade correction 
and morphing parameter and 5\% of PSF to consider for pile-up. The number of 
detector regions to consider independently for pile-up was allowed to change
which led to the best fit value of 38 regions. This is consistent with
the number of 3$\times$3 pixel regions close to excised central core of PSF. 
This approach successfully removed the hard excess due to the pile-up effect.
Finally, we excluded 4.5-7.5 keV energy
range where the line emission is expected to be significant  
and fit  the spectrum again to obtain the fit quality of $\chi^2_{red} = 1.36$.
The result is shown in panel D of Fig. \ref{line_res}.

The final best-fit parameters for the continuum model are following: hydrogen column 
$N_H=2.3\pm0.3\times10^{-21}$ cm$^{-2}$ for {\tt wabs}; seed photons temperature 
$T_0=0.14\pm0.03$ keV, electron temperature $kT=2.27\pm0.06$ keV, optical depth 
 $\tau_p=17.3\pm 0.4$ for {\tt comptt} and {\it bbody}  temperature $T_{bb}=0.87\pm0.01$ keV.
The energy, sigma and equivalent width ($EW$) for {\tt gaussians} used to fit lines are 
$E_L= 1.04\pm 0.02$ keV, $\sigma_L=0.1\pm 0.03$ keV, $EW=27.8\pm9.0$ eV and $E_L=3.19\pm 0.02$ keV, 
$\sigma_L=0.06\pm 0.03$ keV, $EW=4.5\pm2.0$ eV. 
When we  include the iron K$_\alpha$ emission region (4.5-7.5 keV energy band) 
and fit the spectrum with the relativistic and wind line models  
the parameter ranges of the above continuum parameters change insignificantly (see Table 1). 
This indicates that continuum model weakly depends on the assumed line
model.

It is worth noting that the values of the continuum parameters imply
that during this {\it Suzaku} observation  Cyg X-2 was in the ``high/soft'' 
state characterized by  high opacity of geometrically thin configuration. 
 Panel E in Fig. \ref{line_res} shows the residuals of
the  best-fit model with iron region noticed in the data. The apparent 
 line profile is broad and red-skewed.
In fact, the fit with the {\tt gaussian} line model (Model 1)  produces the worst fit quality among 
the applied models  $\chi^2_{red}=1.34$ (see Table 1). 

We apply physically motivated models to describe this red-skewed line profile.
Namely, the relativistic models {\tt diskline} \citep{diskline} and {\tt laor}
\citep{laor} and 
the wind outflow model {\tt windline} (LT07 MC code).
We refer to these models as Model  2, 3 and 4 respectively.
The parameters of the {\tt laor} and {\tt diskline} models are:
the line energy $E_L$, inner and outer disk radius $R_{in}$ and $R_{out}$,
 emissivity index $\beta_L$ and  inclination angle $i_l$.
During fitting these relativistic models to the data the value of $R_{out}$ was fixed at its default value
of 1000 $R_G$  and 400 $R_G$ for {\tt diskline} and the {\tt laor} respectively.
To avoid unphysical results we required the value of the inner radius
$R_{in}$ to be greater than $3 R_G$  as dictated by the NS compactness limit
\citep[see e.g.][]{lat07}. The {\tt windline} model is not a part of the standard set of XSPEC models and
is implemented as a local model during our modeling.
The {\tt windline} model calculates the line profile 
by means of MC simulations.
The main parameters of the model 
are the input line energy $E_L$, the optical depth of the wind $\tau_w$, the wind electron temperature $kT_w$ 
and the dimensionless wind  speed $v/c$. The additional fixed parameter of the model is the 
number of individual photons  $N_{ph}$
to be used in the MC simulation. 

\begin{figure}
\includegraphics[scale=0.7,angle=-90]{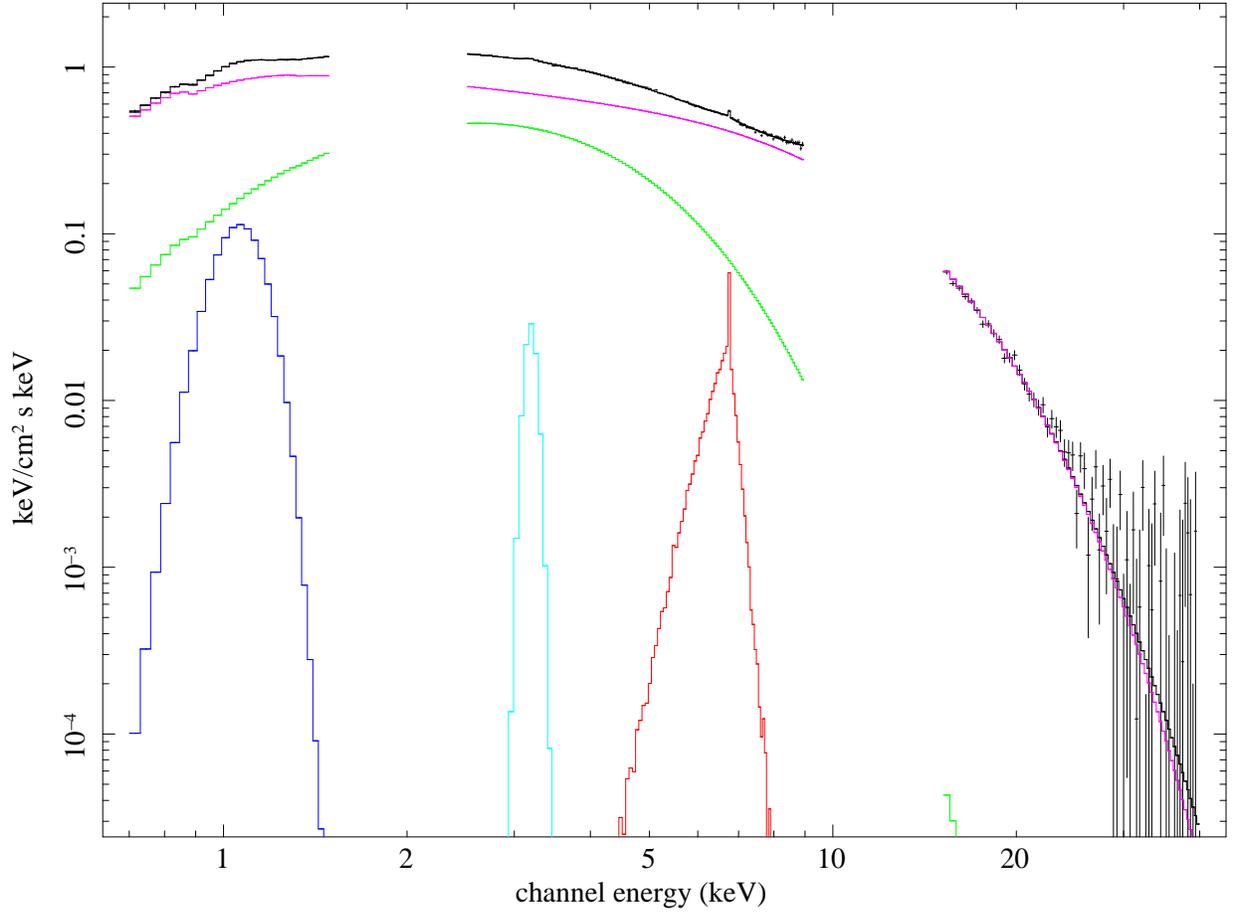}
\caption{Unfolded model fit to the {\it Suzaku} Cyg X-2 spectrum with the Model 4.
The best-fit model of the source spectrum consist of  {\tt comptt} (magenta), {\tt bbody} (green), {\tt windline} (red) and two {\tt gaussians} at 1 keV (blue) and 3.2 keV (turquoise).}
\label{windline_fit}
\end{figure}

We didn't observe any significant dependence of the $\chi^2$-statistic behavior
on the number of photons used in  our line simulations for $N_{ph}$ higher than 3$\times$10$^3$.  
For the presented study we used 5$\times$10$^3$ photons in the {\tt windline} MC 
simulations. We summarize the results of our modeling and fit quality in Table 1. 
The resulting values of $\chi^2_{red}$ for Models 2, 3 and 4 are 1.31, 1.30 and 1.32
respectively. In Figure \ref{windline_fit} we present the unfolded view of spectral fit with 
Model 4 where we used 5$\times$10$^4$ photons in the MC simulation.

\begin{figure}
\includegraphics[scale=0.6,angle=-90]{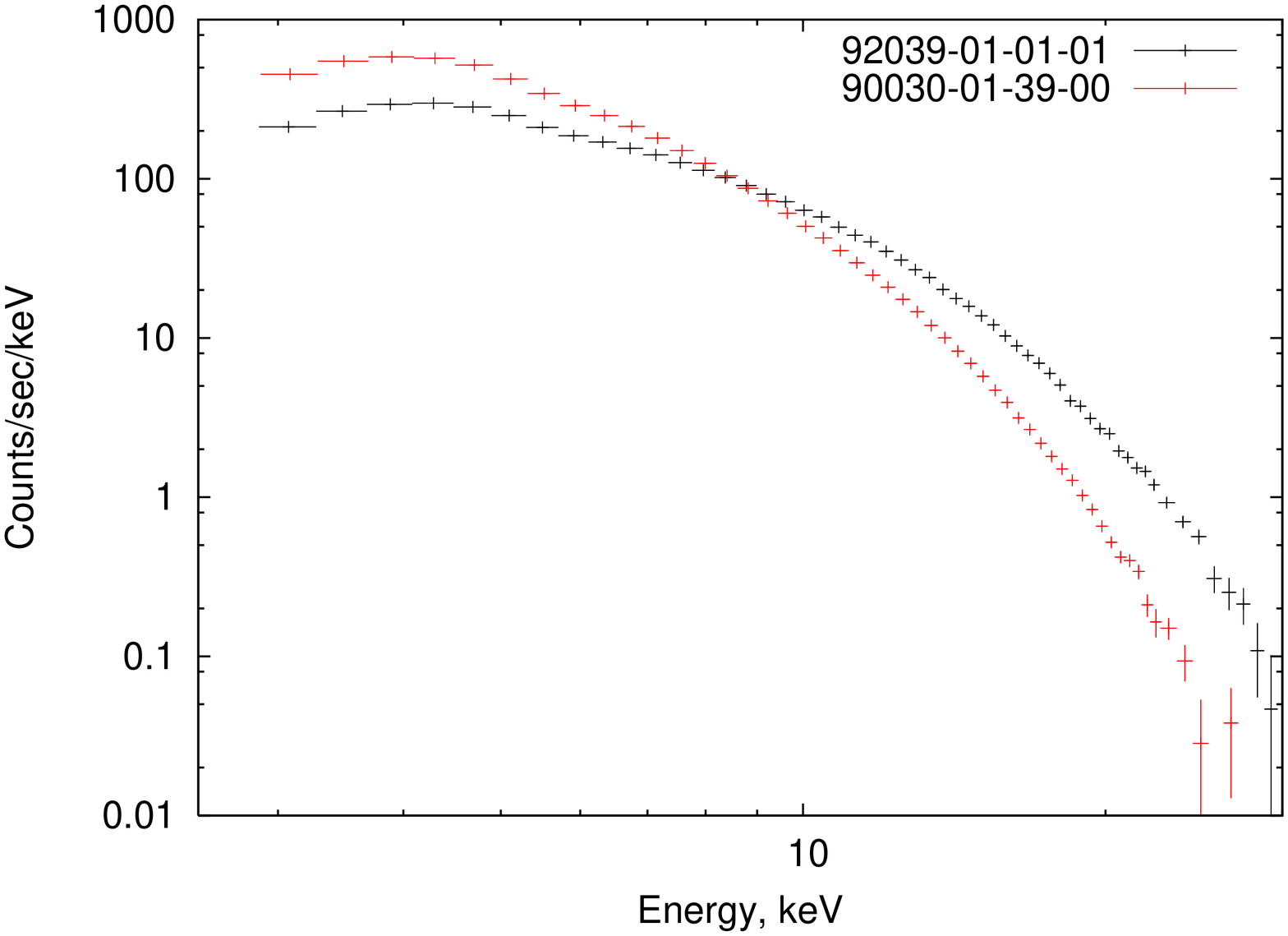}
\includegraphics[scale=0.6,angle=-90]{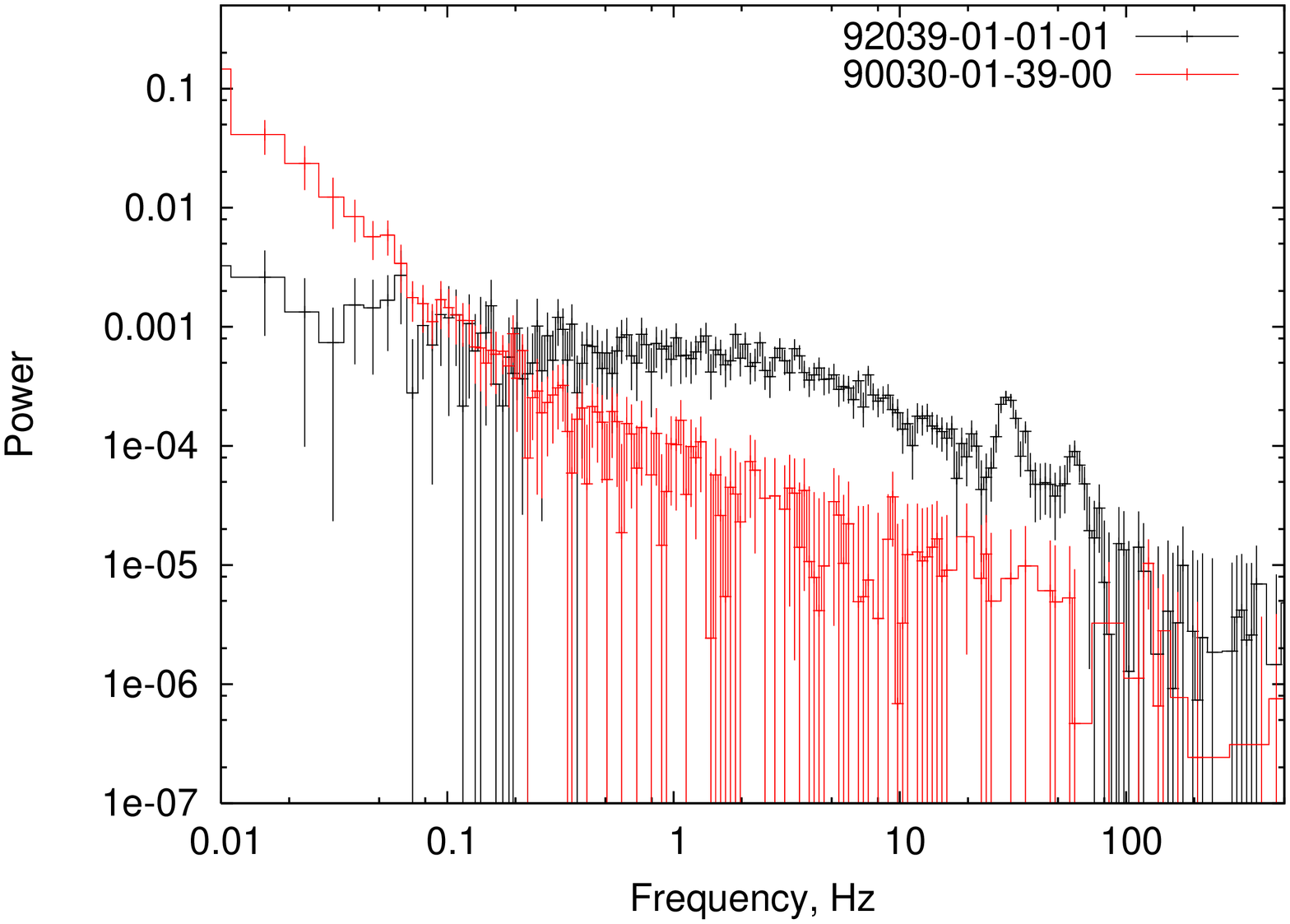}
\caption{ Energy and Power Density spectra as observed by {\it RXTE} during observation 92039-01-01-01 (black) and 
90030-01-39-00 (red) . Spectral characteristics for observation 90030-01-39-00 are similar to those 
measured during {\it Suzaku} observations. Timing analysis shows that high frequency variability during this 
observation is strongly suppressed.}
\label{rxte}
\end{figure}

We also utilize the data from two {\it RXTE} observations made on 
July 25, 2006 and September 4, 2004 (Obs. IDs 92039-01-01-01 and 90030-01-39-00 correspondingly)
to  investigate the evolution of timing properties of Cyg X-2. We extract the {\it RXTE}/PCA energy 
spectra from  Standard2 data modes and we use high resolution modes to calculate power density spectra (PDS). 
We fit {\it RXTE} energy spectra in XSPEC with the model {\tt wabs(comptt+bbody\-+gaussian)}, where {\tt gaussian}
is used to model the iron line. $N_H$ column density was fixed at 
$2.2\times10^{21}$.  We obtain the following best-fit values. For 
ObsID 92039-01-01-01, {\tt comptt}: $T_0=0.001$ keV (unconstrained), $kT_e=3.06\pm0.05$ keV, 
$\tau_p=12.6\pm0.4$, {\tt bbody}: $T_{bb}=1.13\pm0.05$ keV, {\tt gaussian}: 
$E_L=6.64\pm0.17$ keV, $\sigma_L=0.67\pm0.20$ keV, $EW=86\pm25$ eV. For ObsID 90030-01-39-00, 
{\tt comptt}: $T_0=0.16\pm0.06$ keV, $kT_e=2.23\pm0.01$ keV, $\tau_p=14.5\pm3$, 
{\tt bbody}: $T_{in}=1.01\pm0.02$ keV, {\tt gaussian}: $E_L=6.57\pm0.15$ keV, $\sigma_L=0.75\pm0.15$ keV,
$EW=115\pm23$ eV. {\it Spectral parameters for observation ID 90030-01-39-00 match closely those
for the {\it Suzaku} spectrum}. All Sky Monitor (ASM) counts rates in 1.5-3/3-5/5-12 keV energy ranges 
are 14.06/14.85/14.03 cnts/s and 13.8/14.07/13.2 cnts/s  during May 15, 2006 (date of {\it Suzaku} observation)
and during September 4, 2004 (date of 90030-01-39-00 {\it RXTE} observation) respectively, 
showing almost the same fluxes and source hardness. 
This also strongly indicates that during the  {\it Suzaku} observation 401049010 and {\it RXTE} 
observation ID 90030-01-39-00 Cyg X-2 was almost in the same state.
Thus, we can arguably assume that  the timing characteristics  of the source during 
 {\it RXTE} observation 90030-01-39-00 should also be similar to those
  of the  {\it Suzaku} observation.  
In Figure \ref{rxte} we present {\it RXTE} energy spectra along  with the power spectra
 related to {\it RXTE} observations 90030-01-39-00 (red), which are expected to be similar to the analyzed 
{\it Suzaku} observations,  and for the observation 92039-01-01-01 (black).
 
\section{Discussion}

The presented evidence for the red-skewed iron line in Cyg X-2 along with
the detection of the asymmetric lines in Serp X-1, 4U 1820-30,
GX 349+2 \citep[][C08]{bs07} and 4U 1636-536 \citep{pan08}, increases
a number of NS showing this effect to five.
This indicates that
red-skewed line may be as abundant in NS sources as in BHs \citep[e.g.][]{mil07}.
C08 interpreted these observed  line appearances   
as the evidence of relativistic distortion due to the reflection from
inner disk located  close to the NS surface. The main argument in the association
of the line skewness with the inner disk is that the highest observed
kilohertz QPO in these sources is consistent with the Keplerian frequency 
at the inner disk radius predicted by the line profile.

The {\tt diskline} model, when applied to the {\it Suzaku} (Model 2), leads to
the best fit value for the disk inner radius equal to the lower limit
set by the model, i.e $6 GM/c^2$, which formally translates to the radius of $\simeq 12$ km
for the NS mass of 1.4$M_\odot$. 
Model 3, which employs the {\tt laor} component to represent the iron line, 
exhibits the same statistical performance as Model 2 and yields
the radius of 9.5 $R_G$ or 20 km. We should note that
the {\tt laor} model was formulated for the extreme Kerr BH case [see  \citet{laor}]. 
In this model  the spin parameter $j=cJ/GM^2$ was assumed to be close to unity. 
A moderately rotating accreting NS is expected  to have a spin parameter less than 0.5.
The NS spin in Cyg X-2 is not exactly known but if the difference between
kHz QPO of $\sim$364 Hz  is taken as a spin estimate then the space-time background near
Cyg X-2 should satisfy Schwarzshild metric within 10\% margin of error.

C08 found that their values of $R_{in}$ were consistent with the inner disk radius 
predicted by the beat-frequency model
\citep{mlp98} from kHz QPO values. Maximum kHz QPO value observed in
Cyg X-2 is 1007 kHz, which translates to 16.7 km radius if interpreted as
a Keplerian frequency at the inner disk.  This inferred  radius  is  in between the $R_{in}$ values 
obtained using  {\tt diskline} and {\tt laor} models. 
Interpretation of the highest kHz QPO peak as a Keplerian frequency is not unique.
In the framework of the transition layer (TL) model the {\it lower} kHz QPO peak is classified
as a Keplerian frequency \citep[see][and references therein]{t02}. 
The TL model  successfully describes the behavior of kHz QPOs in NS sources.
The TL model QPO classification  leads to the inner disk edge radius of of 22.5 km. 
Beyond this radius the accreting gas enters the TL to adjust its motion to the rotation of the central star.
While the value of the disk inner radius given by the {\tt diskline} model for the iron line
can be considered to be in satisfactory agreement with the beat-frequency QPO model,
it is harder to reconcile with the TL paradigm.

The red-skewed line - kHz QPO connection is based on the {\it highest observed} 
kHz QPO values (C08). However, the duty cycle of kHz QPOs is low. As noted by C08,
 the most compelling evidence
for the inner disk origin of the line would come from the simultaneous observation
of the kHz QPOs  and the red-skewed iron line. At the time of this writing
these two effects were not observed simultaneously from the same source.

The reason for absence of the simultaneous detection of kHz QPOs and broad iron line
may be a lack of the correlated observation of high spectral resolution 
X-ray telescopes (i.e. {\it Suzaku}/XMM-Newton) and {\it RXTE}.
The Cyg X-2 {\it Suzaku} observation analyzed in this Paper is a striking 
example how this lack of simultaneous coverage creates an obstacle in
conducting scientific investigation. Namely, we have to search {\it RXTE}
archive for observation with similar spectral characteristics to estimate
timing properties (see the previous section and the discussion below) while 
simultaneous {\it RXTE} data would allow directly test timing properties
more reliably.

A general idea about the fast timing properties can be inferred
based on the archival {\it RXTE} data by matching the spectral parameters shown by
{\it RXTE} instruments with those observed in the {\it Suzaku} spectrum in question. 
This approach is justified by the
firmly established correlations between spectral characteristics and timing properties
in NS LMXBs \citep{kaa98} and particularly in Cyg X-2 \citep{tks07}.
We searched {\it RXTE} archive for pointing observations nearest to May 16, 2006.
Despite the fact that  {\it RXTE} usually conducts frequent monitoring of Cyg X-2,
we found that {\it Suzaku} observation was made in the middle of 140 day gap in
 {\it RXTE} monitoring of this source. This, and the fact that Cyg X-2 is changing its state on a daily basis,
does not permit us to test subsecond timing variability directly with {\it RXTE}.
Therefore, we have to resort to search for matching spectral properties 
and rely on spectral-variability correlations.
We found that the nearest {\it RXTE} observation with the spectrum 
similar to the one shown by {\it Suzaku} was performed on September 4, 2004 
(ObsID 90030-01-39-00, see the previous section). To compare this  observation 
with the spectral  state characterized by harder spectrum and lower opacity we arbitrarily 
choose the observation 92039-01-01-01 which is fit by {\tt comptt} model with
 parameters  $\tau_p=12.6\pm0.4$ and $kT_e=3.04\pm0.02$. 
In Figure \ref{rxte} we show the energy and power spectra for these observations. 
It is clear that the variability for frequencies higher than 0.1 Hz is strongly suppressed and  a 
``forest''-type PDS  appears at frequencies above 0.1 Hz.
The photon spectrum related to this PDS   
closely matches  the {\it Suzaku} spectral  data.
This  effect of variability  suppression  can be readily explained by a smearing of a 
signal in the opaque wind of optical depth $\tau_{w} $ because the suppression factor is related to
the opacity in the wind as
$\sim e^{-\tau_w}$. A higher {\tt gaussian} equivalent width of 115 eV related to 90030-01-39-00 
{\it RXTE} observation with respect to that of 86 eV during  92086-01-01-01 observation 
 is another factor to support the low variability - strong line connection.

 The broad band variability continuum observed in PDS
can be produced by diffusion of disk small perturbations 
propagating towards central object. These disk  perturbations  result in a modulation of  
inner mass accretion rate, which then 
leads to variability in the X-ray flux \citep{L97}. \citet{tsa07} 
developed   a theory of the diffusive propagation of perturbations in the disk and 
presented a model for the power spectrum formation.
Application of the diffusion theory to the Fourier power spectra of Cyg X-2 and BH source Cyg X-1
led to the conclusion that the fast X-ray variability is produced in two configurations:
in a large cold outer accretion disk and a compact geometrically thick configuration (i.e. transition layer).
Long-term variations of the accretion matter supply at the outer accretion disk 
edge lead to changes of  the source spectral state. Specifically, in  
high-soft state the innermost  region becomes very compact and relatively cold   as a result of  the strong mass accretion.
Presumably, this is
the state which we are dealing with in the case of  {\it RXTE} observation 90030-01-39-00 
and the {\it Suzaku} observation.
However, in this case strong accretion disk presumably
extended close to the central object should produce high frequency perturbations
leading to a broad band PDS continuum or kHz QPOs in the range $\geq 100$ Hz.
None of these is seen in our data. 
This may be attributed to  wind/outflow attenuation.
It could  present a problem for the relativistic disk reflection model of  the  iron emission line
production. We note however, that the origin of fast variability and physical processes governing its evolution
are not yet fully understood. Therefore, the above arguments does not provide a solid proof 
for the wind line model or a dismissal of the relativistic reflection scenario.
Futures studies on the asymmetric emission lines in X-ray binaries 
should address the above points using more substantial
observational data.  Observation of millisecond variability in the same data 
which would require wind opacities of 2$\sim$3 would provide grounds for 
the rejection of the wind model. Such cases have not been observed yet.
On the other hand, further evidence of the correlation of ``forest''-type PDS
with the asymmetric line would rule out  relativistic reflection paradigm.

Cyg X-2 is a Z source \citep{hvdk89} and presumably   accretes 
at a rate close to the Eddington limit. Therefore, strong outflows are expected
in this source, naturally leading to the production of  the red-wing of the iron line
 in this wind/outflow configuration. This hypothesis 
is supported by the fact that Z sources have strong radio counterparts.
\citet{p06} analysis of radio/X-ray correlation in Z and bright atoll
sources that radio and Comptonized emission in these sources originate in
same optically thick plasma with the temperature 2.5-3 keV near NS. 
Moreover, strong requirement for the disk illumination to be
concentrated very close to its inner edge \citep{nan99} in the relativistic line formation
scenario  led  \citet{rb97} to consider  the production of 
fluorescent emission within the innermost stable orbit region  where matter spirals into the compact object.
However,  as indicated by the models of \cite{n00} and \cite{b01},
the ionization of such a disk by the intense X-ray radiation might further invalidate some of 
basic assumptions associated with this interpretation.
These arguments indicate that the wind/outflow paradigm
may indeed be at work as (or at least contribute to) an origin
of the red-skewness in the iron lines in compact sources.
 
\section{Conclusions}

We present the analysis of the {\it Suzaku} spectrum from NS Cyg X-2.
We discover an K$_\alpha$ iron line which shows significant red-wing.
This is the fifth NS source so far to show asymmetric
line profile. We analyze the line in terms of the relativistic emission
from the inner accretion disk and in terms of the wind outflow model.

We conclude that both model are acceptable according to 
the statistical performance. However, the wind model appears to give more adequate explanation which does not
require the accretion disk inner edge to advance close to NS surface. 
We also consider the line production scenarios in the context of the timing
properties. We  identify {\it RXTE} observation which has spectrum very similar to that
during  {\it Suzaku} observation which shows the red-skewed line.
These {\it RXTE} data indicate that the source fast variability is strongly 
suppressed, which can be attributed to smearing in a strong wind. This lack 
of high frequency variability weakens the red-skewed line connection with kHz QPO
and strengthens its connection to a wind/outflow phenomena. 

We acknowledge the valuable suggestions by anonymous referee which considerably improved
this Paper.






\begin{deluxetable}{lllll}
\tablewidth{0pt}
\tabletypesize{\footnotesize}
\tablecaption{Best-fit Parameters for the {\it Suzaku} Spectrum of Cygnus X-2 }
\tablehead{
\colhead{Parameter} &
\colhead{Model 1\tablenotemark{a}} &
\colhead{Model 2\tablenotemark{b}} &
\colhead{Model 3\tablenotemark{c}} &
\colhead{Model 4\tablenotemark{d}}}
\startdata
$N_H$, cm$^2$ ......... & 0.23$\pm$0.03 & 0.23$\pm$0.03 & 0.23$\pm$0.03 & 0.23$\pm$0.03 \\
$T_{bb}$, keV ...... & 0.87$\pm$0.01 & 0.87$\pm$0.01 & 0.87$\pm$0.01 & 0.87$\pm$0.01 \\
$T_0$, keV ........... & 0.14$\pm$0.03 & 0.14$\pm$0.04 & 0.14$\pm$0.03 & 0.14$\pm$0.03 \\
$kT_e$, keV ......... & 2.28$\pm$0.02 & 2.32$\pm$0.02 & 2.35$\pm$0.02 & 2.28$\pm$0.02 \\
$\tau_p$ .................... & 17.2$\pm$0.4 & 17.1$\pm$0.4 & 17.0$\pm$0.4 & 17.2$\pm$0.4 \\
$E_L$, keV ..........& 6.65$\pm$0.03 & 6.64$\pm$0.14 & 6.49$\pm$0.10 & 6.78$\pm$0.04 \\
$\sigma$, keV .............& 0.22$\pm$0.04 & - & - & - \\
$\beta_L$ ...................& - & 2.3$\pm$0.5 & 3.0\tablenotemark{f} & - \\
$R_{in}$, $R_G$ ..........& - & 6.0\tablenotemark{e} & 9.5$\pm$1.3 & - \\  
$i_L$, deg .............& - & 30.2$\pm$6.8 & 34.0$\pm$4.1 & - \\
$\tau_w$.....................& - & - & - & 1.29$\pm$0.35 \\
$v/c$, $10^{-2}$..........& - & - & - & 1.43$\pm$0.14 \\
$kT_w$, keV .........& - & - & - & 0.16$\pm$0.10 \\
$EW_L$, eV.......... & 28.5$\pm$7.1& 43.2$\pm$11.0 & 50.6$\pm$13.0 & 37$\pm$15 \\
$\chi^2/N_{dof}$............& 251.2/187 & 243.3/185 & 241.4/186 & 243.5/185 \\
\enddata

\tablenotetext{a}{XSPEC Model: {\tt wabs(comptt+bbody+gaussian)}}

\tablenotetext{b}{XSPEC Model: {\tt wabs(comptt+bbody+diskline)}}

\tablenotetext{c}{XSPEC Model: {\tt wabs(comptt+bbody+laor)}}
\tablenotetext{d}{XSPEC Model: {\tt wabs(comptt+bbody+windline)}}
\tablenotetext{e}{Parameter pegged into the lower boundary value set by the model}

\tablenotetext{f}{Parameter is fixed. Thawing of the parameter leads to unphysical results.}

\label{pars_tab} 
\end{deluxetable}

\newpage

\newpage

\newpage

\end{document}